\documentclass[twocolumn,trackchanges]{aastex62}
\usepackage{graphicx}

\def\xmm {\emph{XMM--Newton}}
\def\flux {\mbox{erg cm$^{-2}$ s$^{-1}$}}
\def\lum {\mbox{erg s$^{-1}$}}
\def\src {\mbox{3X\,J0042}}
\graphicspath{{./}{figures/}}

\submitjournal{ApJL on April 28, 2018.}

\shorttitle{Discovery of a 3\,s X-Ray Pulsar in M\,31}
\shortauthors{Rodr\'iguez Castillo et al.}

\begin{document}

\title{Discovery of 3\,s pulsations from the Brightest Hard X-ray Source in M\,31}

\correspondingauthor{Guillermo Rodr\'iguez}
\email{guillermo.rodriguez@oa-roma.inaf.it, grodriguezcas@gmail.com}

\author[0000-0003-3952-7291]{Guillermo A. Rodr\'iguez Castillo}
\affil{INAF--Osservatorio Astronomico di Roma, via Frascati 33, I-00078 Monteporzio Catone, Italy}

\author[0000-0001-5480-6438]{Gian Luca Israel}
\affiliation{INAF--Osservatorio Astronomico di Roma, via Frascati 33, I-00078 Monteporzio Catone, Italy}

\author[0000-0003-4849-5092]{Paolo Esposito}
\affiliation{Anton Pannekoek Institute for Astronomy, University of Amsterdam, Science Park 904,
1098\,XH Amsterdam, The Netherlands}

\author[0000-0001-6289-7413]{Alessandro Papitto}
\affiliation{INAF--Osservatorio Astronomico di Roma, via Frascati 33, I-00078 Monteporzio Catone, Italy}


\author[0000-0002-0018-1687]{Luigi Stella}
\affiliation{INAF--Osservatorio Astronomico di Roma, via Frascati 33, I-00078 Monteporzio Catone, Italy} 

\author[0000-0002-6038-1090]{Andrea Tiengo}
\affiliation{Scuola Universitaria Superiore IUSS, piazza della Vittoria 15, I-27100 Pavia, Italy}
\affiliation{INAF--Istituto di Astrofisica Spaziale e Fisica Cosmica - Milano, via E. Bassini 15, I-20133 Milano, Italy}
\affiliation{INFN--Istituto Nazionale di Fisica Nucleare, Sezione di Pavia, via A. Bassi 6, I-27100 Pavia, Italy}

\author[0000-0001-6739-687X]{Andrea De Luca}
\affiliation{INAF--Istituto di Astrofisica Spaziale e Fisica Cosmica - Milano, via E. Bassini 15, I-20133 Milano, Italy}

\begin{abstract}

We report the discovery with \xmm\ of 3-s X-ray pulsations from 3XMM\,J004232.1+411314, a dipping source that dominates the hard X-ray emission of M\,31. This finding unambiguously assesses the neutron star (NS) nature of the compact object.  We also measured an orbital modulation of 4.15\,h and a projected semi-axis at $a_{\mathrm{X}} \sin i= 0.6$\,lt-s, which implies a low-mass companion of about 0.2--0.3\,$M_{\odot}$ assuming a NS  of 1.5\,$M_{\odot}$ and an orbital inclination $i=70\degr$--80$\degr$. The barycentric orbit-corrected pulse period decreased by $\sim$28\,ms in about 16 yr, corresponding to an average spin-up rate of $\dot{P} \sim -6 \times 10^{-11}$\,s\,s$^{-1}$ ;  pulse period variations, probably caused to by X-ray luminosity changes, were observed on shorter time scales. 
We identify two possible scenarios for the source: a mildly  magnetic NS with  $B_{\mathrm{p}}\simeq$ few $\times10^{10}$\,G if the pulsar is far from its equilibrium period $P_{\mathrm{eq}}$, and a relatively young highly magnetic NS with $B_{\mathrm{eq}}\simeq 10^{13}$\,G if spinning close to $P_{\mathrm{eq}}$. 

\end{abstract}

\section{Introduction} \label{sec:intro}

At a distance $d= 780$~kpc \citep{holland98}, M\,31 is the closest major galaxy to the Milky Way. Despite a number of extensive studies, only one accreting X-ray neutron star (NS) pulsar binary has been securely identified so far, 3XMM\,J004301.4+413017 (spin and orbital periods of 1.2\,s and 1.3\,d, respectively; \citealt{esposito16}). 
Here we report the discovery of pulsations from 
3XMM\,J004232.1+411314 (\src), a luminous X-ray source located in the inner bulge of M\,31, about 3.7\,arcmin from the center, which was detected in all the X-ray surveys of M\,31 conducted with good-resolution imaging instruments \citep[see][and references therein]{vulic16}. Its luminosity was found to be consistently above $10^{37}$\,\lum. Based on its spectral shape, luminosity, and moderate variability, the source was suggested to be an X-ray binary. Using \emph{NuSTAR} \citet{yukita17} pinpointed \src\ as the soft X-ray counterpart of Swift\,J0042.6+4112, whose hard X-ray emission ($\gtrsim$20\,keV) dominates the bulge of M\,31 (\citealt{baumgartner13,revnivtsev14}). After examining all available multi-wavelength information, \citet{yukita17} concluded that 
\src\ could be either an X-ray pulsar with an intermediate-mass companion (stellar counterparts heavier than 3\,$M_\odot$ are disfavored by the \emph{HST} images) or a symbiotic X-ray binary. 
During the analysis of the \xmm\ X-ray data archive carried out within the EXTraS project,\footnote{Exploring the X-ray Transient and variable Sky, see \citet{deluca16} and \url{http://www.extras-fp7.eu}.} \citet{marelli17} discovered dips recurring with a period of $\sim$4\,h in the light curve of \src, likely reflecting the orbital period of the system. This prompted them to propose that the source is a low-mass X-ray binary (LMXB) seen at high inclination ($60\degr$--80$\degr$).
In this Letter, we report the discovery of 3-s pulsations in the \xmm\ data of \src; pulse arrival time delays provide  an orbital period of 4.15\,h, which confirms the suggestion by \citet{marelli17}. We discuss the nature of the source in the light of these results.

\begin{figure*}
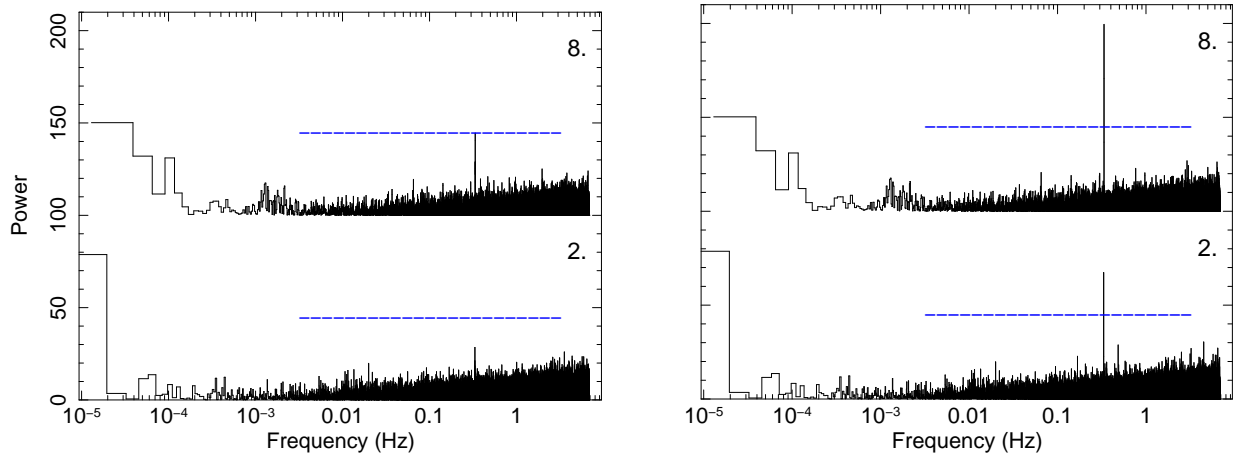

\centering
{\includegraphics[width=6cm,angle=-90]{xronos_nocorr_all.eps}\includegraphics[width=6cm,angle=-90]{xronos_corr_all.eps}}
\caption{Left panel: Power spectral density (PSD) of the pn 0.2--12~keV original light curves of two out of the nine data sets in which pulsations were detected (the labels identify the observations as in Table\,\ref{tab:detections}; the upper PSDs were shifted upward by 100 for displaying purposes). The blue dashed lines indicate the 3.5$\sigma$ detection threshold, which takes into account the number of Fourier frequencies in the PSD. Right panel: same as before, but after correcting the photon arrival times of the two data sets for the orbital parameters derived from the phase fitting of Obs.\,\#9 ($P_{\mathrm{orb}}$ and $a_{\mathrm{X}}\sin i$) and from the orbital correction pipeline (local $T_{\mathrm{node}}$; see Sect.\,\ref{sec:datanres} for details).\label{fig:dps}}
\end{figure*}

\section{Data Analysis and Results} \label{sec:datanres}

The region of \src\ was observed 48 times by \xmm\ with the EPIC pn and MOS cameras operating in different observational modes (the most relevant differences being the time resolution and size of the field of view; see \citealt{struder01,turner01}).  
The raw observation data files (ODF) were processed with the Science Analysis Software (SAS) v.16 and subsequent versions. Time intervals with high particle background were filtered out. Photon event lists were extracted from a 31 arcsec radius around the source, while the background was estimated from
a nearby circular region with the same radius, which excluded other sources and CCD gaps. Our timing analysis was based on the pn data, since they offer the highest time resolution. Photon arrival times were converted to the barycentre of the Solar system with the SAS task BARYCEN.

\begin{figure*}
\centering
\resizebox{\hsize}{!}{\includegraphics[angle=-90]{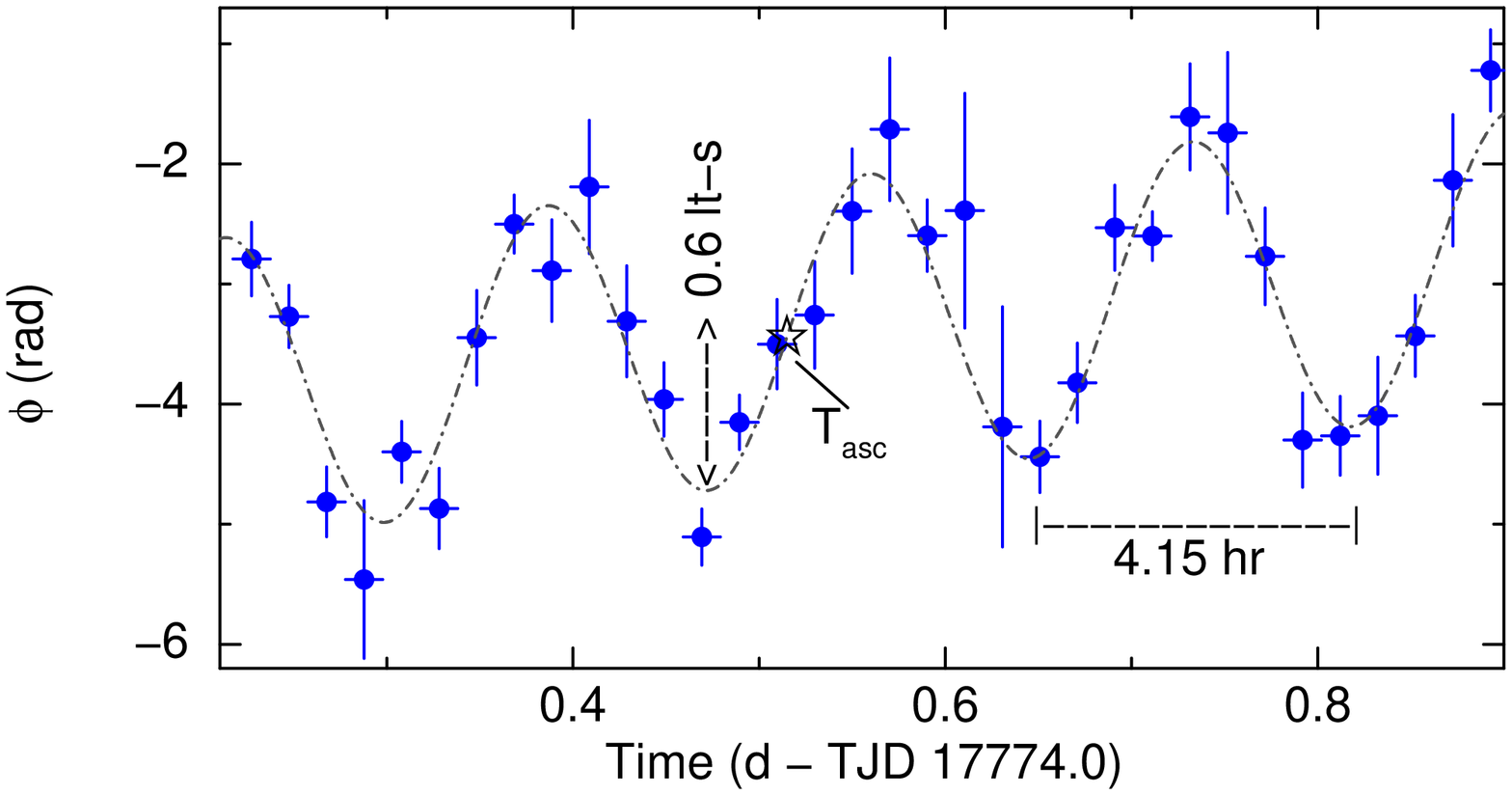} 
\includegraphics[width=11.5cm,angle=-90]{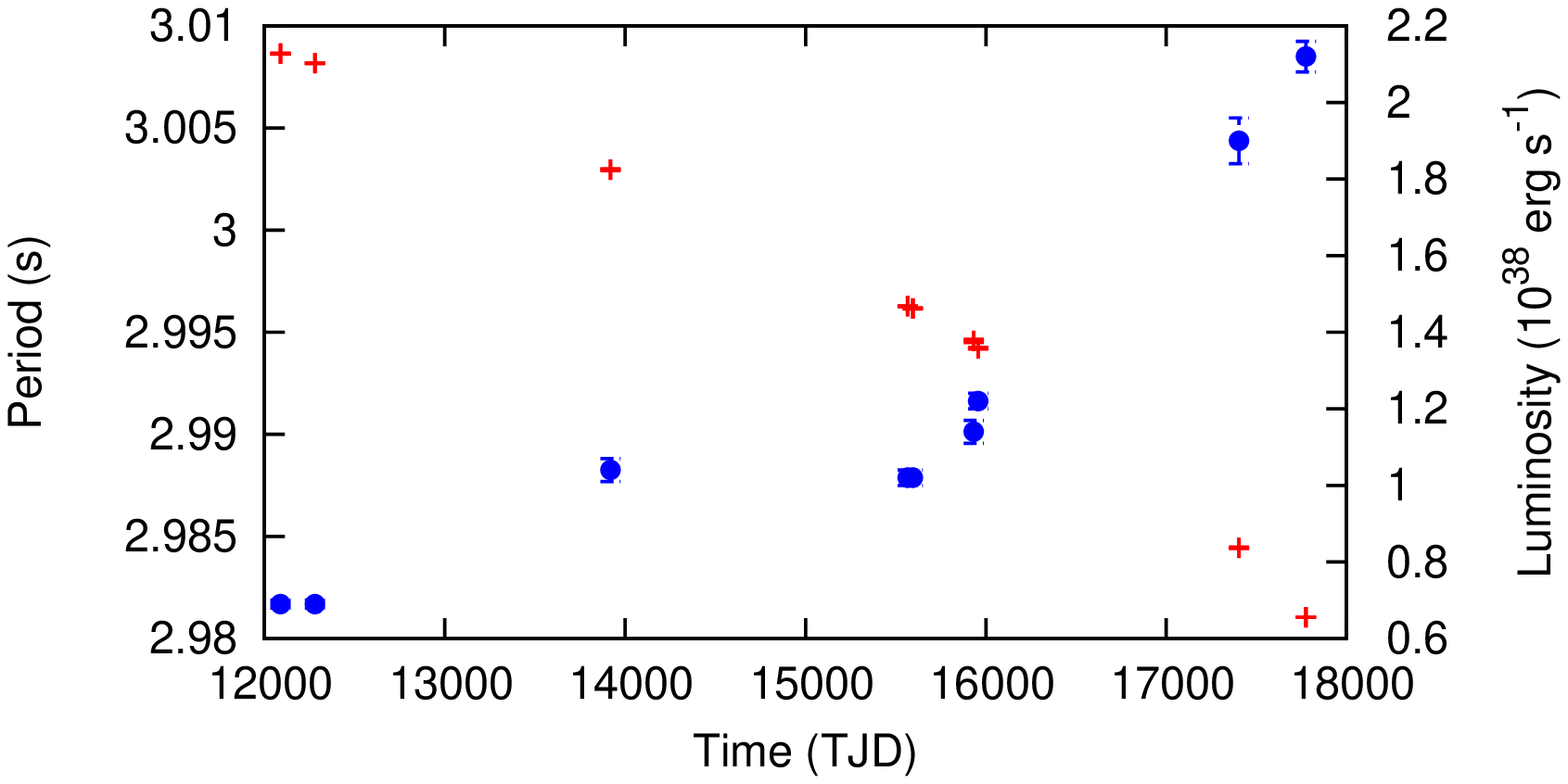}}
\caption{Left panel: Pulse phases as a function of time for the longest \xmm\ observation of \src\ (Obs.\,\#9). The Doppler orbital modulation is evident and the main orbital parameters measured ($P_{\mathrm{orb}}$,  $a_{\mathrm{X}}\sin i$, and $T_{\mathrm{node}}$) are indicated graphically. 
Right panel: The period (red crosses) and 0.2--12\,keV luminosity evolution (blue dots). The spin-up rate increase occurring approximately after TJD 16000 appears to correlate with the luminosity increase. 
\label{fig:timing}}
\end{figure*}

The technique we used to search the pn data for periodic signals is a generalization of the Fourier-based procedure outlined in \citet{israel96}, where we also included "de-acceleration" corrections for both the first period derivative and the Doppler delays. 
Each photon arrival time is corrected according to the expression:  $t' = t - (a_{\mathrm{X}}\sin i/c) \sin[2\pi(t - T_{\mathrm{node}})/P_{\mathrm{orb}}]$, where $t$ is the photon arrival time, $a_{\mathrm{X}}\sin i$ is the semi-axis projection on the orbital plane, $c$ is the speed of light, $P_{\mathrm{orb}}$ is the orbital period, and $T_{\mathrm{node}}$ is the time of the ascendant node. This correction is applied for each combination of $P_{\mathrm{orb}}$, $a_{\mathrm{X}}\sin i$, and $T_{\mathrm{node}}$ in a 3-dimension grid with given parameter steps. Then, the procedure is reiterated within a vector of $\dot{P}$ values. Since this approach is CPU-time-consuming, we adopted the orbital period of $\sim$4.01\,h inferred from the dips recurrence by \citet{marelli17} as our initial guess and took into account also the (reasonable) constraints on the maximum mass of the companion derived by \citet{yukita17}. 

\begin{deluxetable}{lcccccc}
\tablecaption{The \xmm\ observations in which the 3-s spin period was detected. Figures in parentheses represent the 1$\sigma$ uncertainties in the least significant digit. \label{tab:detections}}
\tablewidth{0pt}
\tablehead{
\colhead{Data set} & \colhead{Date\tablenotemark{a}} & \colhead{Period} &
\colhead{PF\tablenotemark{b}} & \colhead{$L_{\mathrm{X}}$\tablenotemark{c}} & \nocolhead{Flux} \\
\colhead{\#~~ Obs.ID} & \colhead{(TJD)} & \colhead{(s)} &
\colhead{(\%)} & \colhead{(\lum)} & \nocolhead{($10^{12}\,$\flux)}
}
\decimals
\startdata
1. 0109270101 & 12089.6 & 3.00864(2)  & 11(2) & 0.69(1) \\
2. 0112570101 & 12281.1 & 3.00816(1)  & 14(2) & 0.69(1) \\
3. 0405320501 & 13918.7 & 3.00295(5)  & 14(2) & 1.04(3) \\
4. 0650560301 & 15565.9 & 2.99627(2)  & 11(2) & 1.02(2) \\
5. 0650560601 & 15596.1 & 2.99617(3)  & 14(2) & 1.02(2) \\
6. 0674210301 & 15933.2 & 2.99457(8)  & 11(2) & 1.14(3) \\
7. 0674210601 & 15957.2 & 2.99422(4)  & 11(2) & 1.22(2) \\
8. 0764030301 & 17403.9 & 2.98443(3)  & 16(1) & 1.90(6) \\
9. 0790830101 & 17774.6 & 2.981041(6) & 14(1) & 2.12(4) \\
\enddata
\tablenotetext{\tablenotemark{a}}{Mid-point of the observation.}
\tablenotetext{\tablenotemark{b}}{Pulsed fraction, defined as the semi-amplitude of the sinusoid divided by the average source count rate.}
\tablenotetext{\tablenotemark{c}}{In units of $10^{38}$ and in the 0.2--12\,keV band.}
\end{deluxetable}

\begin{figure}
\centering
{\includegraphics[angle=-90,scale=0.65]{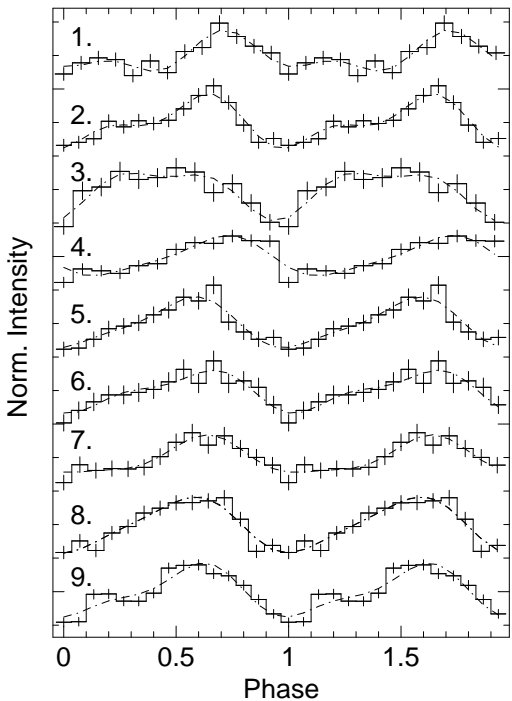}}
\caption{\xmm/pn orbit-corrected  0.2--12\,keV light curves folded at the  period measured in each data set where the $\sim$3\,s signal was   detected. Two cycles are shown for clarity, and the curves have been  aligned in phase and shifted vertically in an arbitrary fashion for displaying purposes.\label{fig:pulses}}
\end{figure}

The first run of the search over the observation with the longest exposure and the highest count rate (Obs.ID 0790830101, see Table\,\ref{tab:detections}) revealed a strong signal (12$\sigma$ confidence level, c.l.) at approximately 2.98\,s for the grid point [$P_{\mathrm{orb}}=14280$\,s, $a_{\mathrm{X}}\sin i=0.54$\,lt-s, $T_{\mathrm{node}}=17774.3$\, truncated Julian day (TJD)]. We performed a phase-fitting analysis of the $\sim$4 orbital cycles covered by this observation (see \citealt{dallosso03} for details), in order to measure more precisely the orbital parameters. The phase of the pulses was determined over 34 consecutive time intervals, as shown in Fig.\,\ref{fig:timing} (left panel). The best-fitting orbital parameters are reported in Table\,\ref{tab:orbit}. These honed values were then fed as inputs to the search algorithm for the remaining 47 observations. We detected the $\sim$3-s signal at a c.l. higher than 3.5$\sigma$ in eight more observations (see Table\,\ref{tab:detections} and Fig.\,\ref{fig:dps}; note that in obs.ID 0109270101 the signal was unambiguous but just below the detection threshold). For each detection, the period estimate was refined with the phase-fitting analysis. 

Figure\,\ref{fig:timing} (right panel) shows the period evolution,  
whereas Fig.\,\ref{fig:pulses} displays a collection of all light curves folded at the best-fit period of each data set (see also Table\,\ref{tab:detections}). No significant change of the pulsed fraction as a function of energy was detected, likely owing to the softness of the source and poor statistics at energies above $\sim$4\,keV. 
From TJD 12089 (2001 Jun 29) to 17774 (2017 Jan 21), the average spin-up rate of \src\ was $\dot{P} \sim-5\times10^{-11}$\,s\,s$^{-1}$. However, the period evolution displays clear departures from a constant $\dot{P}$.
Both the luminosity and spin evolution of the source (see right panel of Fig.\,\ref{fig:timing} and figure\,1 of \citealt{marelli17}) appear to have changed around TJD 16000, after which  a substantial steepening of the  spin-up rate took place and the luminosity approximately doubled. We estimate that the average $\dot{P}$ was  $\sim$$-4.2\times10^{-11}$ and $-8.5 \times 10^{-11}$\,s\,s$^{-1}$ before and after TJD $\sim$16000, respectively (note that the period measurements are too sparse to constrain better the epoch and the pace of the $\dot{P}$ change).

For all remaining datasets, we derived 3$\sigma$ upper limits on the pulsed fraction over an interval of trial periods determined by using  the $\dot{P}$ measurements. The average 3$\sigma$ upper limit on the pulse fraction in the 39 observations is $\sim$26\%, the most stringent limit being 18\%. It is therefore possible that the flux modulation at 3\,s is always present (the highest value measured, that of Obs.ID 0764030301, is 16\%, cf. Table\,\ref{tab:detections}), but remained undetected in those data sets owing to poor statistics.

\section{Discussion} \label{sec:diss}




The accretion disk around a magnetized neutron star is truncated at a radius which is usually estimated as a fraction $\xi$ of the Alfv\'en radius:
\begin{equation}
 R_{\mathrm{in}}=68.5\,\xi_{0.5}\,M_{*,1.4}^{1/7}\,R_{*,10}^{-2/7}\,(\eta L_{38})^{-2/7} \mu_{28}^{4/7}~\mathrm{km}.
\label{eq:rin} 
\end{equation}
Here,  $M_{*,1.4}$ and $R_{*,10}$ are the NS mass and radius in units of 1.4\,$M_{\odot}$, and $10$\,km, respectively, $L_{38}$ is the X-ray luminosity in units of $10^{38}$\,\lum, $\eta=L_{\mathrm{X}} R_*/(GM\dot{M})$=1 is the accretion efficiency, $\mu_{28}$ is the magnetic dipole moment in units of $10^{28}$\,G\,cm$^2$, and $\xi$ is a factor that depends on the details of the disk-magnetosphere interaction and ranges between 0.5 and 1 (see, e.g. \citealt{bozzo09} for a discussion of the various models; \citealt{campana18} recently determined $\xi =0.5$ by using data from different classes of accreting magnetic stars undergoing transitions to the propeller regime). The disk must be truncated at a radius larger than the NS radius, $R_*$, and smaller than the co-rotation radius ($R_{\mathrm{co}}=3485\,M_{*,1.4}^{1/3}$\,km, for a 3-s pulsar) for accretion-powered X--ray pulsations to be produced. The former limit originates from the requirement that the NS magnetosphere is not buried under the NS surface, the latter implies that the NS rotation does not inhibit accretion through the propeller effect. Inserting the observed X-ray luminosity range ($L_{38}=0.5$--$2$) in Eq.\,\ref{eq:rin}, the condition $R_*<R_{\mathrm{in}}<R_{\mathrm{co}}$ is satisfied for $\mu_{28}=0.05$--$700$. The observation of X-ray pulsations then binds the magnetic field strength at the poles of the neutron star to a broad range of $B_{\mathrm{p}}=2\mu R_*^{-3}=10^9$--$1.4\times10^{13}\,R_{*,10}^{-3}$\,G, where the largest values correspond to a disk truncated at the co-rotation radius.

The torque applied by the disk onto the NS is expressed by $N=-2\pi I \dot{P} / P^2 =\dot{M}\sqrt{G M_* R_{\mathrm{in}}} F(\omega)$, where $I$ is the NS moment of inertia, $\omega=(R_{\mathrm{in}}/R_{\mathrm{co}})^{3/2}$ is the fastness parameter, and the function $F(\omega)$ describes the reversal of the torque when the inner disk radius approaches the co-rotation radius \citep{gl79}. 
Assuming that \src\ is a {\it slow rotator} spinning far from equilibrium and thus that the disk is truncated well inside the co-rotation radius (i.e. $\omega\ll1$ and $F(\omega)=1$), and using  Eq.\,\ref{eq:rin} to express the inner disk radius, we get
\begin{equation}
\dot{P}_{-11}=2.75\,I_{45}^{-1}\,(\eta L_{38})^{6/7} \xi_{0.5}^{1/2}\,M_{*,1.4}^{-3/7}\,R_{*,10}^{6/7}\,\mu_{28}^{2/7}
\label{eq:pdot}\nonumber
\end{equation}
for a $P=3$\,s pulsar.
Here, $\dot{P}_{-11}$ is the spin-up rate in units of $10^{-11}$ and $I_{45}$ is the NS moment of inertia in units of $10^{45}$\,g\,cm$^{-2}$. Considering an average luminosity of $L_{38}\simeq2$, the average spin-up rate observed from \src, $\dot{P}_{-11}\sim -6$, is explained if the pulsar has a dipole magnetic moment of $\mu_{28}\simeq 2.7\,I_{45}^{7/2}\,M_{*,1.4}^{3/2}\,R_{*,10}^{3}\,\eta^{-5/2}\,\xi_{0.5}^{-7/4}$. Considering the range of values that can be taken by $\xi=0.5$--1, this corresponds to a magnetic field at the NS poles of $B_{\mathrm{p}}\simeq(1.5$--$5)\times10^{10}$\,G. The spin down rate of \src\ was observed to increase from $\dot{P}_{-11}\sim -4.2$ to $-8.5$ while the X-ray luminosity increased from $L_{38}=0.9$ to 2.0, suggesting a relation $P\propto L_{\mathrm{X}}^\alpha$ with $\alpha\simeq 1$. This value is close to that predicted by simple models of the  disk-magnetosphere interaction ($\alpha=6/7$) when the pulsar is a {\it slow rotator}, lending 
support to the above interpretation. However, the spin-up time-scale $P/\dot{P}\simeq1600$~yr is very short compared to the typical duration of the spin-up accretion phase of pulsars in LMXBs, $10^8$--$10^9$\,yr (e.g. \citealt{tauris12}), casting doubts on this interpretation. On the other hand, there are known cases of X-ray pulsars accreting close to or above the Eddington luminosity with much shorter spin-up time-scales than expected (e.g. some Ultraluminous X-ray Pulsars; \citealt{israel17}). In this scenario a related object to \src might be IGR\,J17480--2446, a 90~ms transient X-ray pulsar in a LMXB with a companion with mass $>$0.4\,$M_{\odot}$, which belongs to the globular cluster Terzan\,5 \citep{papitto11} and an inferred  magnetic field in the (0.5--$1.5)\times10^{10}$\,G range \citep[][see also \citealt{cavecchi11}]{papitto12}.
\begin{deluxetable}{lr}
\tablecaption{Orbital parameters of \src.  Numbers in parentheses represent 1$\sigma$ uncertainties on the last significant digit.
\label{tab:orbit}}
\tablewidth{0pt}
\tablehead{
\colhead{Parameter} & \colhead{Value} 
}
\startdata
Orbital period, $P_{\mathrm{orb}}$ (h) & $4.15(4)$\\
Epoch of ascending node, $T_{\rm node}$ (TJD) & $17774.525(3)$ \\
Projected semi-axis, $a_{\mathrm{X}}\sin i$ (lt-s) & $0.59(4)$ \\
Mass function, $f(M)$ ($M_{\sun}$) & $7(2) \times 10^{-3}$ \\
Companion mass\tablenotemark{a} ($M_{\sun}$) & 0.22--0.32 \\
\enddata
\tablenotetext{\tablenotemark{a}}{For $i$ between $70$ and $80\degr$
, as estimated by \cite{marelli17}, 
and a NS with mass of 1.5\,$M_{\odot}$. The range widens to 0.19--0.46\,$M_{\odot}$ for NS masses in the (1--2.8)\,$M_{\sun}$ interval. }
\end{deluxetable}

Alternatively, the pulsar might be spinning close to the equilibrium value (i.e. $R_{\mathrm{in}}\simeq R_{\mathrm{co}}$). The magnetic field required  in this case would be $B_{\mathrm{eq}}=1.4\times10^{13}\,R_{*,10}^{-5/6}\,M_{*,1.4}^{1/3}\,\xi^{-7/4}\,(\eta L_{38})^{1/2}$\,G, 
i.e. much larger than that of the previous scenario. The expected spin-up torque depends crucially on the shape of the function $F(\omega)$. Taking $F(\omega)=1-\omega^2$ \citep{ertan09}, the observed average spin up rate is reproduced with $\omega=0.92$ (i.e., $R_{\mathrm{in}}=0.95 R_{\mathrm{co}}$). If \src\ is spinning close to equilibrium, it is expected that small variations of the mass accretion rate might induce large variation of the spin-up rate or even torque reversals (see \citealt{camero10} for the case of 4U\,1626--67). In this scenario, the spin-up timescale would not necessarily be related to the time elapsed since the onset 
of accretion and the hard X-ray spectral component of \src\ extending up to tens of keV might be more easily interpreted. 

Our results demonstrate that \src\ is a 3\,s X-ray pulsar 
accreting close to the Eddington rate from a low mass companion. 
The evolution of the spin period together with the other 
source properties indicate that the pulsar might be either in the slow rotator
regime and possess a magnetic field of $\sim$$10^{10.5}$~G or be a  
a quasi-equilibrium rotator with a field of $\sim$$10^{13}$~G. 
In the latter interpretation, a transition to the propeller regime and thus a 
drastic reduction of the source luminosity would ensue in response 
to a decrease of the disk mass inflow rate. Therefore, the detection of a transient 
behavior from \src\ would provide decisive support in favor of the 
quasi-equilibrium rotator interpretation.


\acknowledgments
This research is based on observations obtained with \xmm, an ESA science mission with instruments and contributions directly funded by ESA Member States and NASA. This research was supported by high performance computing resources  awarded by CINECA (MARCONI), under the ISCRA initiative to the ``PASTA-X project'', and also through the INAF--CHIPP high performance computing project resources and support. PE acknowledges funding in the framework of the NWO Vidi award A.2320.0076. LS, AP and GLI acknowledge financial contribution from ASI-INAF agreement I/037/12/0. AP also acknowledges funding from the EU Horizon 2020 Framework Programme for Research and Innovation under the Marie Sk\l{}odowska-Curie Individual Fellowship grant agreement 660657-TMSP-H2020-MSCA-IF-2014.


\bibliographystyle{aasjournal}
\bibliography{biblio}

\begin{thebibliography}{}
\expandafter\ifx\csname natexlab\endcsname\relax\def\natexlab#1{#1}\fi

\bibitem[{{Baumgartner} {et~al.}(2013){Baumgartner}, {Tueller}, {Markwardt},
  {Skinner}, {Barthelmy}, {Mushotzky}, {Evans}, \& {Gehrels}}]{baumgartner13}
{Baumgartner}, W.~H., {Tueller}, J., {Markwardt}, C.~B., {et~al.} 2013, \apjs,
  207, 19

\bibitem[{{Bozzo} {et~al.}(2009){Bozzo}, {Stella}, {Vietri}, \&
  {Ghosh}}]{bozzo09}
{Bozzo}, E., {Stella}, L., {Vietri}, M., \& {Ghosh}, P. 2009, \aap, 493, 809

\bibitem[{{Camero-Arranz} {et~al.}(2010){Camero-Arranz}, {Finger}, {Ikhsanov},
  {Wilson-Hodge}, \& {Beklen}}]{camero10}
{Camero-Arranz}, A., {Finger}, M.~H., {Ikhsanov}, N.~R., {Wilson-Hodge}, C.~A.,
  \& {Beklen}, E. 2010, \apj, 708, 1500

\bibitem[{{Campana} {et~al.}(2018){Campana}, {Stella}, {Mereghetti}, \& {de
  Martino}}]{campana18}
{Campana}, S., {Stella}, L., {Mereghetti}, S., \& {de Martino}, D. 2018, \aap,
  610, A46

\bibitem[{{Cavecchi} {et~al.}(2011){Cavecchi}, {Patruno}, {Haskell}, {Watts},
  {Levin}, {Linares}, {Altamirano}, {Wijnands}, \& {van der Klis}}]{cavecchi11}
{Cavecchi}, Y., {Patruno}, A., {Haskell}, B., {et~al.} 2011, \apjl, 740, L8

\bibitem[{{Dall'Osso} {et~al.}(2003){Dall'Osso}, {Israel}, {Stella},
  {Possenti}, \& {Perozzi}}]{dallosso03}
{Dall'Osso}, S., {Israel}, G.~L., {Stella}, L., {Possenti}, A., \& {Perozzi},
  E. 2003, \apj, 599, 485

\bibitem[{{De Luca} {et~al.}(2016){De Luca}, {Salvaterra}, {Tiengo},
  {D'Agostino}, {Watson}, {Haberl}, \& {Wilms}}]{deluca16}
{De Luca}, A., {Salvaterra}, R., {Tiengo}, A., {et~al.} 2016, in Astrophysics
  and Space Science Proceedings, Vol.~42, The Universe of Digital Sky Surveys,
  ed. N.~{Napolitano}, G.~{Longo}, M.~{Marconi}, M.~{Paolillo}, \& E.~{Iodice}
  (Springer International Publishing, Cham), 291

\bibitem[{{Ertan} {et~al.}(2009){Ertan}, {Ek{\c s}i}, {Erkut}, \&
  {Alpar}}]{ertan09}
{Ertan}, {\"U}., {Ek{\c s}i}, K.~Y., {Erkut}, M.~H., \& {Alpar}, M.~A. 2009,
  \apj, 702, 1309

\bibitem[{{Esposito} {et~al.}(2016){Esposito}, {Israel}, {Belfiore}, {Novara},
  {Sidoli}, {Rodr{\'{\i}}guez Castillo}, {De Luca}, {Tiengo}, {Haberl},
  {Salvaterra}, {Read}, {Salvetti}, {Sandrelli}, {Marelli}, {Wilms}, \&
  {D'Agostino}}]{esposito16}
{Esposito}, P., {Israel}, G.~L., {Belfiore}, A., {et~al.} 2016, \mnras, 457, L5

\bibitem[{{Ghosh} \& {Lamb}(1979)}]{gl79}
{Ghosh}, P., \& {Lamb}, F.~K. 1979, \apj, 234, 296

\bibitem[{{Holland}(1998)}]{holland98}
{Holland}, S. 1998, \aj, 115, 1916

\bibitem[{{Israel} \& {Stella}(1996)}]{israel96}
{Israel}, G.~L., \& {Stella}, L. 1996, \apj, 468, 369

\bibitem[{{Israel} {et~al.}(2017){Israel}, {Belfiore}, {Stella}, {Esposito},
  {Casella}, {De Luca}, {Marelli}, {Papitto}, {Perri}, {Puccetti}, {Castillo},
  {Salvetti}, {Tiengo}, {Zampieri}, {D'Agostino}, {Greiner}, {Haberl},
  {Novara}, {Salvaterra}, {Turolla}, {Watson}, {Wilms}, \& {Wolter}}]{israel17}
{Israel}, G.~L., {Belfiore}, A., {Stella}, L., {et~al.} 2017, Science, 355, 817

\bibitem[{{Marelli} {et~al.}(2017){Marelli}, {Tiengo}, {De Luca}, {Salvetti},
  {Saronni}, {Sidoli}, {Paizis}, {Salvaterra}, {Belfiore}, {Israel}, {Haberl},
  \& {D'Agostino}}]{marelli17}
{Marelli}, M., {Tiengo}, A., {De Luca}, A., {et~al.} 2017, \apjl, 851, L27

\bibitem[{{Papitto} {et~al.}(2011){Papitto}, {D'A{\`i}}, {Motta}, {Riggio},
  {Burderi}, {di Salvo}, {Belloni}, \& {Iaria}}]{papitto11}
{Papitto}, A., {D'A{\`i}}, A., {Motta}, S., {et~al.} 2011, \aap, 526, L3

\bibitem[{{Papitto} {et~al.}(2012){Papitto}, {Di Salvo}, {Burderi}, {Belloni},
  {Stella}, {Bozzo}, {D'A{\`i}}, {Ferrigno}, {Iaria}, {Motta}, {Riggio}, \&
  {Tramacere}}]{papitto12}
{Papitto}, A., {Di Salvo}, T., {Burderi}, L., {et~al.} 2012, \mnras, 423, 1178

\bibitem[{{Revnivtsev} {et~al.}(2014){Revnivtsev}, {Sunyaev}, {Krivonos},
  {Tsygankov}, \& {Molkov}}]{revnivtsev14}
{Revnivtsev}, M.~G., {Sunyaev}, R.~A., {Krivonos}, R.~A., {Tsygankov}, S.~S.,
  \& {Molkov}, S.~V. 2014, Astronomy Letters, 40, 22

\bibitem[{{Str{\"u}der} {et~al.}(2001){Str{\"u}der}, {Briel}, {Dennerl},
  {Hartmann}, {Kendziorra}, {Meidinger}, {Pfeffermann}, {Reppin}, {Aschenbach},
  {Bornemann}, {Br{\"a}uninger}, {Burkert}, {Elender}, {Freyberg}, {Haberl},
  {Hartner}, {Heuschmann}, {Hippmann}, {Kastelic}, {Kemmer}, {Kettenring},
  {Kink}, {Krause}, {M{\"u}ller}, {Oppitz}, {Pietsch}, {Popp}, {Predehl},
  {Read}, {Stephan}, {St{\"o}tter}, {Tr{\"u}mper}, {Holl}, {Kemmer}, {Soltau},
  {St{\"o}tter}, {Weber}, {Weichert}, {von Zanthier}, {Carathanassis}, {Lutz},
  {Richter}, {Solc}, {B{\"o}ttcher}, {Kuster}, {Staubert}, {Abbey}, {Holland},
  {Turner}, {Balasini}, {Bignami}, {La Palombara}, {Villa}, {Buttler},
  {Gianini}, {Lain{\'e}}, {Lumb}, \& {Dhez}}]{struder01}
{Str{\"u}der}, L., {Briel}, U., {Dennerl}, K., {et~al.} 2001, \aap, 365, L18

\bibitem[{{Tauris} {et~al.}(2012){Tauris}, {Langer}, \& {Kramer}}]{tauris12}
{Tauris}, T.~M., {Langer}, N., \& {Kramer}, M. 2012, \mnras, 425, 1601

\bibitem[{{Turner} {et~al.}(2001){Turner}, {Abbey}, {Arnaud}, {Balasini},
  {Barbera}, {Belsole}, {Bennie}, {Bernard}, {Bignami}, {Boer}, {Briel},
  {Butler}, {Cara}, {Chabaud}, {Cole}, {Collura}, {Conte}, {Cros}, {Denby},
  {Dhez}, {Di Coco}, {Dowson}, {Ferrando}, {Ghizzardi}, {Gianotti}, {Goodall},
  {Gretton}, {Griffiths}, {Hainaut}, {Hochedez}, {Holland}, {Jourdain},
  {Kendziorra}, {Lagostina}, {Laine}, {La Palombara}, {Lortholary}, {Lumb},
  {Marty}, {Molendi}, {Pigot}, {Poindron}, {Pounds}, {Reeves}, {Reppin},
  {Rothenflug}, {Salvetat}, {Sauvageot}, {Schmitt}, {Sembay}, {Short},
  {Spragg}, {Stephen}, {Str{\"u}der}, {Tiengo}, {Trifoglio}, {Tr{\"u}mper},
  {Vercellone}, {Vigroux}, {Villa}, {Ward}, {Whitehead}, \& {Zonca}}]{turner01}
{Turner}, M.~J.~L., {Abbey}, A., {Arnaud}, M., {et~al.} 2001, \aap, 365, L27

\bibitem[{{Vulic} {et~al.}(2016){Vulic}, {Gallagher}, \& {Barmby}}]{vulic16}
{Vulic}, N., {Gallagher}, S.~C., \& {Barmby}, P. 2016, \mnras, 461, 3443

\bibitem[{{Yukita} {et~al.}(2017){Yukita}, {Ptak}, {Hornschemeier}, {Wik},
  {Maccarone}, {Pottschmidt}, {Zezas}, {Antoniou}, {Ballhausen}, {Lehmer},
  {Lien}, {Williams}, {Baganoff}, {Boyd}, {Enoto}, {Kennea}, {Page}, \&
  {Choi}}]{yukita17}
{Yukita}, M., {Ptak}, A., {Hornschemeier}, A.~E., {et~al.} 2017, \apj, 838, 47

\end{thebibliography}

\end{document}